\documentclass[12pt,epsf]{article}
\usepackage{graphics,epsf}
\begin{document}
\title{Non-radial null geodesics in spherical dust collapse}
\author{Filipe C Mena$^\natural$ and Brien C Nolan$^\flat$\\
{\small $^\natural$ School of Mathematical Sciences,}
\\{\small Queen Mary, University of London, London E1 4NS, UK.}
\\{\small $^\flat$ School of Mathematical Sciences, }\\{\small Dublin City
University, Glasnevin, Dublin 9, Ireland.}}

\maketitle
\begin{abstract}The issue of the local visibility of the
shell-focussing singularity in marginally bound spherical dust
collapse is considered from the point of view of the existence of
future-directed null geodesics with angular momentum which emanate
from the singularity. The initial data (i.e. the initial density
profile) at the onset of collapse is taken to be of class $C^3$.
Simple necessary and sufficient conditions for the existence of a
naked singularity are derived in terms of the data. It is shown
that there exist future-directed non-radial null geodesics
emanating from the singularity if and only if there exist
future-directed radial null geodesics emanating from the
singularity. This result can be interpreted as indicating the
robustness of previous results on radial geodesics, with respect
to the presence of angular momentum.
\newline Pacs:04.20.Dw, 04.20.Ex
\end{abstract}
\newtheorem{assump}{Assumption}
\newtheorem{theorem}{Theorem}
\newtheorem{prop}{Proposition}
\newtheorem{corr}{Corollary}
\newtheorem{lemma}{Lemma}
\newcommand{\mo}{\mu_0}
\newcommand{\mot}{\frac{\mo}{3}}
\section{Introduction}
It has long been known that naked singularities may arise in the
gravitational collapse of inhomogeneous dust spheres; see
\cite{mzh}, where the existence of naked so-called shell-crossing
singularities was first demonstrated. The physical significance of
these singularities has been questioned, primarily on the basis
that they are gravitationally weak according to the definition of
Tipler \cite{newman,nolan99}. However, it remains to be shown how
to obtain a unique evolution to the future of a shell-crossing
singularity. A more disturbing type of naked singularity was
discovered and christened the shell-focussing singularity in
\cite{eardleysmarr}. This singularity was further studied in
\cite{christo} and \cite{newman}, where the first results
regarding the role of regular initial data in determining the
causal nature of the singularity were derived. This question has
been studied extensively since then (see \cite{jj-israel} and
references therein). The main conclusion of these studies is that
for a given initial mass distribution, the remaining initial data
(namely the velocity distribution of the dust cloud) may be chosen
so that the collapse results in either a black hole or a naked
singularity \cite{dj97}. Furthermore, there are open subsets (of
the appropriate initial data space) of choices for the velocity
distribution which lead to these conclusions. In this sense, both
censored and naked singularities are stable
\cite{sary-ghate,mena-tav}. Indications of the instability of the
Cauchy horizon associated with the naked singularity have arisen
in perturbative studies \cite{IHN}.

The aim of this study is to address the stability of the naked
singularity from another point of view. All results to date on the
visibility of the singularity relate to the existence of
future-pointing {\em radial} causal geodesics emanating from the
singularity. (In all but one case \cite{deshingkar}, where radial
{\em time-like} geodesics are treated, the results relate solely
to radial {\em null} geodesics.) Our goal is to investigate the
occurrence or otherwise of {\em non-radial} null geodesics
emanating from the singularity. This is of importance as
non-radial geodesics provide a better model of physically
realistic trajectories than do radial geodesics. On the
mathematical side, radial null geodesics passing through a point
$p$ of space-time constitute a set of measure zero in the set of
null geodesics through $p$ and so must be considered to be
extremely specialized. Furthermore, investigating non-radial
geodesics gives an indication of how angular momentum (at the
level of the metric) might influence the system.

The structure of this paper is as follows. In Section 2, we
discuss the field equations and occurrence of singularities and
give a clear definition of the initial data which are being
studied. This is essential for a rigorous study of the possible
outcomes. The initial data consist of just one function $\mu(r)$
of a single variable on an interval $[0,b]$, where $b$ is the
initial radius of the dust cloud. We take $\mu\in C^3[0,b]$; this
is more general than the $C^\infty$ or analytic data often
studied. In Section 3, we derive necessary and sufficient
conditions, in terms of the initial data, for the existence of
radial null geodesics emanating from the singularity. Our results
agree with those of previous studies \cite{jj-israel,dj97,JJS96},
but have the two main advantages of (i) overcoming some potential
mathematical deficiencies of previous approaches, for example in
the case of the proof of necessary conditions for the occurrence
of naked singularities; (ii) giving a complete decomposition of
the $C^3$ initial data space into regions leading to naked and
censored singularities. Also, the analysis used here is quite
different from the approach of previous authors and sheds some new
light on the nature of the singularity. We study non-radial null
geodesics in Section 4. Our main conclusion is this: there exist
non-radial null geodesics emanating from the singularity if and
only if there exist radial null geodesics emanating from the
singularity. Some concluding comments are given in Section 5. All
our considerations are restricted to the local visibility of the
singularity. A bullet $\bullet$ indicates the end of a proof. We
use $8\pi G=c=1$.
\section{Field equations and singularities}
We study spherical inhomogeneous dust collapse for the marginally
bound case. The line element is \cite{Lemaitre,Tolman,Bondi}
\begin{equation}
ds^2 = -dt^2 +(R_{,r}dr)^2 +R^2(r,t)d\Omega^2,\label{eq1}
\end{equation}
where $d\Omega^2$ is the line element for the unit 2-sphere. We use subscripts to denote partial derivatives. The
field equations yield
\begin{eqnarray}
R_{,t}&=&-\sqrt{\frac{m(r)}{R}},\label{feq1}\\
\rho &=& \frac{m_{,r}}{R^2R_{,r}},\label{feq2}
\end{eqnarray}
the latter equation defining the density $\rho(r,t)$ of the dust;
$m(r)$ is arbitrary\footnote{$m(r)$ will, in fact, be taken to
satisfy certain differentiability conditions and inequalities
defined ahead in the text.} . ($m$ will be referred to as the
mass, although in fact $m$ equals twice the Misner-Sharp mass. In
particular, this locates the apparent horizon at $R=m$.) Solving
(\ref{feq1}) gives
\begin{equation}
R^3(r,t)=\frac{9}{4}m(t_c(r)-t)^2,\label{req}
\end{equation}
where $t_c(r)$ is another arbitrary function which, in the process of collapse,
corresponds to the time of arrival of each shell $r$ to the singularity.

\begin{assump}
The collapse proceeds from a regular initial state; i.e. at time
$t=0$, (i) there are no singularities (all curvature invariants are
finite) and (ii) there are no trapped surfaces:
\begin{equation}
m(r)<R(r,0).
\end{equation}
\end{assump}
We have the freedom of a coordinate rescaling in (\ref{eq1}) which we can use to set $R(r,0)=r$. This gives
\begin{equation} t_c(r) = \frac{2}{3}\sqrt{\frac{r^3}{m}}.\label{feq3}
\end{equation}
Notice that this leaves just one arbitrary function, $m(r)$.
There is a curvature singularity called the shell focussing singularity along $t=t_c(r)$,
so the ranges of the coordinates $r,t$ are $0\leq r<\infty$ and $0\leq t<t_c(r)$.
Assuming a dust sphere of finite radius,
we can restrict the range of $r$ to $[0,b]$ for some $b>0$, and match it to a Schwarzchild exterior.
During the collapse of the dust sphere
there can be also a curvature singularity given by $R_{,r}=0$ along $t=t_{sc}(r)$, where
\begin{equation}
t_{sc}(r) =2\frac{\sqrt{rm}}{m_{,r}}.
\end{equation}
It is known that this so-called shell-crossing singularity is
gravitationally weak \cite{newman,nolan99}, though what this means
in terms of continuability of the geometry is not yet known. Since we are primarily
interested in the shell focussing singularity we impose:

\begin{assump}
Along each world-line $r=$constant, the shell-crossing singularity does not precede
the shell-focussing singularity. That is, $t_c(r)<t_{sc}(r)$ for all $r>0$.
\end{assump}
This is equivalent to taking $R_{,r}>0$ for all $r>0$, and yields
\begin{equation}
m_{,r}<3\frac{m}{r},\qquad r>0.
\label{nosc}
\end{equation}
We note that a no-shell crossing condition should imply $t'_c(r)>0$ and, in this case,
this is in fact equivalent to (\ref{nosc}).

In order to obtain rigorous results relating to the initial data,
we need to be very clear about what those initial data are.
This can be phrased in terms of the class of functions to which $m(r)$ may belong.
The freely specifiable initial data consists of just one function: the initial density
\begin{equation}
\mu(r) :=\rho(r,0) =\frac{m_{,r}}{r^2}.
\end{equation}
Through the next assumptions we ensure the positivity of the
matter-density $\rho$ during the collapse. We impose a certain
differentiability level on $\mu$ which is higher than one would
normally like to assume ($C^1$), but allows for a comprehensive
discussion of censorship.

\begin{assump}
\begin{enumerate}
\item The mass function $m(r)$ is strictly increasing on $[0,b]$.
\item The density function $\mu (r)$ is such that
\begin{equation}
\mu\in C^3[0,b],\qquad \mo:=\mu(0)\neq 0.\label{mu}\end{equation}
\end{enumerate}
\end{assump} We define three functions $m_i, i=1,2,3$ as
follows. The definitions are meant to apply for whatever values of
$r$ the quantities exist.
\begin{eqnarray}
m(r)&=:&r^3(\mot+m_1),\qquad m_1(0)=0;\label{m1}\\
m_{,r}(r)&=:&r^2(\mo+m_2),\label{m2}\\
m_{,rr}(r)&=:&r(2\mo+m_3),\qquad m_3(0)=0.\label{m3}
\end{eqnarray}
Notice that $m_2(r)=\mu(r)-\mo$, and so $m_2\in C^3[0,b]$ and
$m_2(0)=0$.

We now turn to the question of how the no-shell
crossing condition restricts the initial data $m(r)$. In order to
do that we prove a number of results which constrain both $m(r)$ and
its derivatives.

\begin{prop}
The no-shell crossing condition (\ref{nosc}) is equivalent to
\begin{equation}
m_2(r)<3m_1(r),  r\in(0,b],\label{nosc1}
\end{equation}
which is equivalent to
\begin{equation}
m_1^\prime(r)<0, r\in (0,b].\label{nosc2}
\end{equation}
\end{prop}

\noindent{\bf Proof:} This is immediate from the definitions
above$\bullet$

\begin{corr}
The no-shell crossing condition yields $m_2<3m_1<0$ on $(0,b]$.
\end{corr}

It is important to
study in more detail the properties of $m(r)$ at $r=0$ since they will play an important role
in deciding whether a null geodesic
can escape from the singularity.
We will now demonstrate some conditions on the values of the derivatives
of $m(r)$ at $r=0$. The proof of the next result
makes use of elementary calculus to describe some important
relationships between the derivatives of the functions $m_i(r)$,
$i=1,2,3$ which will be of use in the following section.
\begin{prop}
\label{hello}
$m_1,m_3 \in C^3[0,b]$ and
\begin{eqnarray}
m_1^\prime(0)&=&\frac14 m_2^\prime(0),\qquad m_3^\prime(0)=3m_2^\prime(0);\label{p1eq1}\\
m_1^{\prime\prime}(0)&=&\frac15 m_2^{\prime\prime}(0),\qquad m_3^{\prime\prime}(0)=4m_2^{\prime\prime}(0);\label{p1eq2}\\
m_1^{\prime\prime\prime}(0)&=&\frac16
m_2^{\prime\prime\prime}(0),\qquad
m_3^{\prime\prime\prime}(0)=5m_2^{\prime\prime\prime}(0).\label{p1eq3}
\end{eqnarray}
\end{prop}

\noindent{\bf Proof:} Clearly, it is only differentiability at the origin which needs to be checked.
By the definitions above, we can write
\begin{eqnarray}
m_1(r)=\frac{1}{r^3}\int_0^rx^2m_2(x)\,dx,\label{m1int}\\
m_2(r)=\frac{1}{r^2}\int_0^rxm_3(x)\,dx.\label{m2int}
\end{eqnarray}
Using Taylor's theorem to first order, we can write
\begin{equation}
m_2(x) = xm_2^\prime({\hat x}),
\end{equation}
where the mapping $x\mapsto {\hat x}(x):[0,b]\to[0,x]$ is continuous. Then
\begin{eqnarray*}
m_1(r)&=&\frac{1}{r^3}\int_0^r x^3m_2^\prime({\hat x})\,dx\\
&=&\frac{m_2^\prime({\bar r})}{r^3}\int_0^r x^3\,dx\\
&=&\frac{r}{4}m_2^\prime({\bar r}),
\end{eqnarray*}
where in the second line we used the generalised mean value
theorem for integrals. The mapping $r\mapsto {\bar
r}(r):[0,b]\to[0,r]$ is continuous. Then
\begin{eqnarray*}
\lim_{r\to 0^+}\frac{m_1(r)-m_1(0)}{r}&=&
\lim_{r\to 0^+}\frac{m_1(r)}{r}\\
&=&\lim_{r\to 0^+}\frac14 m_2^\prime({\bar r})\\
&=&\lim_{r\to 0^+}\frac14 m_2^\prime(0),
\end{eqnarray*}
which says that $m_1^\prime(0)$ exists and equals the quantity asserted.
The results on the higher derivatives of $m_1$ are obtained in a similar manner.
Differentiating (\ref{m2int}) and rearranging, we can write
\begin{eqnarray} \frac{m_3(r)}{r}=m_2^\prime(r)+2\frac{m_2(r)}{r}.\label{m2p}\end{eqnarray}
then taking the limit and using l'Hopital's rule for the second
term on the right hand side shows that $m_3^\prime(0)$ exists and
equals the quantity asserted. Results for the higher derivatives
of $m_3$ are obtained by analysing the derivatives of (\ref{m2p})
$\bullet$

As a corollary of the above results one can write useful conditions on the derivatives of the
functions $m_2$ such that the no-shell crossing condition is satisfied.

\begin{corr} The no-shell crossing condition yields:
\begin{enumerate}
\item
$m_2^\prime(0)\leq 0$.
\item
If $m_2^\prime(0)=0$, then $m_2^{\prime\prime}(0)\leq 0$.
\item
If $m_2^\prime(0)=m_2^{\prime\prime}(0)=0$, then $m_2^{\prime\prime\prime}(0)\leq 0$.
\end{enumerate}
\end{corr}
To summarise the main points of this section,
we have proved a number of results which establish conditions
on the mass function such that the process of
collapse starts from a regular initial state and
the shell-crossing of the dust spheres is avoided thus ensuring that the collapse will
end in a (central) shell-focussing singularity. In what follows we will
consider the question of
whether this singularity is naked or covered.
\section{Radial null geodesics}
\begin{figure}[!htb]
\centerline{\def\epsfsize#1#2{0.6
#1}\epsffile{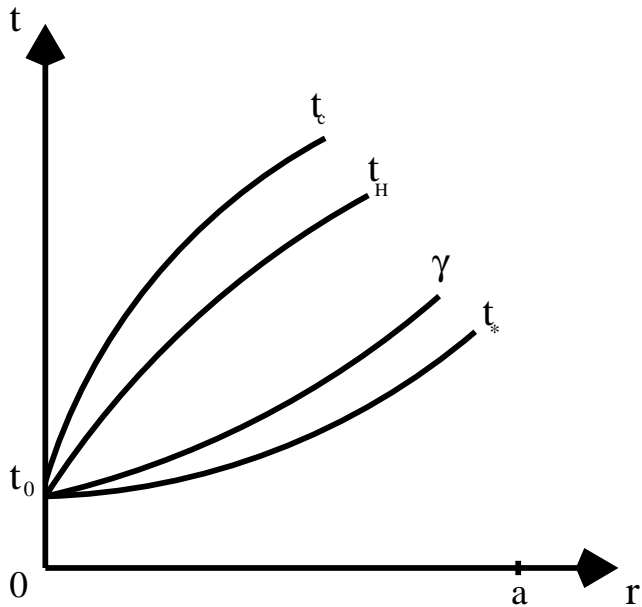}}
\caption{\label{figure1}
Schematic diagram of the singularity structure in the radial $2-$space. In the region
$\Omega=\{(r,t):r<a,t_{*}<t<t_H\}$ no outgoing radial null geodesic $\gamma$ can cross
either $t_H$ or $t_*$ ($t_q$ in the final case), as $r$ decreases,
 and so must extend back to the singularity
$(r=0,t=t_0)$. If $t_H$ is decreasing in a neighbourhood of the
origin, then the singularity must be censored.}
\end{figure}

We recall that a singularity is called {\em locally naked} if in
its neighbourhood there are future-directed causal geodesics which
emanate from the singularity. In this section we study the case of
radial null geodesics\footnote{In fact the results of this section
apply to arbitrary outgoing radial null curves; an affine
parameter is not introduced at any stage.}. Using (\ref{eq1}),
(\ref{req}) and (\ref{feq3}) one can write the equation governing
the outgoing radial null geodesics as
\begin{equation}
\frac{dt}{dr}=R_{,r}=\frac{1}{2}m^{-\frac{2}{3}}
\left(\sqrt{\frac{r^3}{m}}-\frac32 t\right)^{-\frac{1}{3}}
\left(2\sqrt{rm}-m_{,r}t\right).
\label{rngeq}
\end{equation}
We know that if a radial null geodesic (RNG) emanates from the shell focussing singularity,
then it must immediately move into the regular (untrapped) region.
That is, it must precede the apparent horizon, which is given by $R=m$, i.e. by
\begin{equation}
t=t_H(r) = \frac23(\frac{m}{r^3})^{-1/2}-\frac23 m>0.
\label{thdef}
\end{equation}
Furthermore, the only `point' of the shell focussing singularity
which may be visible is $r=0$, whereat
$t_c(0)=\frac23(\mot)^{-1/2}=:t_0$. So we are seeking RNG's that
extend back to $(r=0,t=t_0)$. Henceforth, we shall refer to this
point as `the singularity' which will be called {\em naked} if it
is, at least, locally naked. We shall now derive sufficient
conditions on $m(r)$ such that the singularity is naked. The main
idea of the proof is that in order to find geodesics $\gamma$
escaping from the singularity one must ensure that they lie below
a curve given by $t=t_H(r)$ (i.e. they precede the time of the
apparent horizon) and above another curve $t=t_*(r)$ which ensures
$R_{,rt}>0$ as well as the no-shell crossing condition. This idea
is represented schematically in Figure \ref{figure1}.

\begin{prop}
If $m_2^\prime(0)\neq 0$, or if $m_2^\prime(0)=0$ and $m_2^{\prime\prime}(0)\neq 0$, then the singularity is naked. More precisely, defining
\begin{equation}
t_*(r)=(\frac{m}{r^3})^{-1/2}-\frac{\sqrt{rm}}{m_{,r}}>0,\label{tsdef}
\end{equation}
then there exists $c>0$ such that every outgoing radial null geodesic which
passes through the region
\[ \Omega_{naked}:=\{ (r,t): 0<r<a, t_*(r)<t<t_H(r) \}\]
originates at the singularity.
\end{prop}

\noindent{\bf Proof:}
First, we note that $t_H>t_*$ if and only if
\[ m_1 >\frac13 m_2+\frac23 r^3(\mot+m_1)^{3/2}(\mo+m_2),\]
and so subject to the hypotheses of the Proposition and the
conditions on the mass, there exists a sufficiently small $c$ so
that the region $\Omega_{naked}$ is non-empty. Note also that
$t_H(0)=t_*(0)=t_0$. It is easily established that $t_H(r),t_*(r)$
are differentiable on $[0,b]$, and we find
\begin{equation}
t_*^\prime(r)=-\frac{1}{18}m_2^\prime(0)(\mot)^{-3/2}-\frac{1}{45}m_2^{\prime\prime}(0)(\mot)^{-3/2}r
+ O(r^2). \label{tsdev}
\end{equation}
Now let $p$ be any point of $\Omega_{naked}$. Then there exists a
unique solution of (\ref{rngeq}) through $p$, which for $r>0$ is a
differentiable curve $t_{rad}(r)$. Decreasing $r$ along $t_{rad}$,
the RNG must either cross $t=t_H$, $t=t_*$ or run into the
singularity. Using (\ref{rngeq}), we find that
\begin{equation}
t_{rad}^\prime|_{t=t_H}= t_H^\prime + m_{,r}\geq t_H^\prime\qquad \mbox{\rm{for all}}~r\in[0,b],
\end{equation}
with equality holding only at the origin. Thus in the direction of increasing $r$, $t_{rad}$
crosses $t_H$ from below,
and so extending {\em back } to $r=0$, we see that $t_{rad}$ must stay below $t_H$.
We can also calculate that
\begin{equation}
t_{rad}^\prime|_{t=t_*}=2^{-2/3}(\mot+m_1)^{-1}(\mo+m_2)^{1/3}(3m_1-m_2)^{2/3}.
\end{equation}
It is then a straightforward matter to check that in each of the two cases referred to in the hypotheses, there exists $d>0$ such that, for all $r\in(0,d)$,
\[ t_{rad}^\prime|_{t=t_*}< t_*^\prime(r).\]
Thus in the direction of increasing $r$, $t_{rad}$ crosses $t_*$ from above, and so extending {\em back } to $r=0$, we see that $t_{rad}$ must stay above $t_*$.

Hence we see that the geodesic $t_{rad}$ must extend back to the
singularity $(r=0,t=t_0)$ $\bullet$

We note that one has $t>t_*(r)$ if and only if $R_{,rt}>0$. This will be of
importance when we turn to the study of non-radial null geodesics.
We also note that the condition $t_*>0$ gives
\begin{equation}
m_{,r}>\frac{m}{r}.
\end{equation}
In the last proposition we have shown sufficient conditions for
nakedness. We will now show necessary conditions on $m(r)$ such
that the singularity is censored and which help us to isolate in a
{\em final case} the remaining possible naked solutions. As in the
previous proposition the proof follows by comparing the slopes of
the curves depicted in Figure \ref{figure1}.
\begin{prop}
If $m_2^\prime(0)=m_2^{\prime\prime}(0)=0$ and $|m_2^{\prime\prime\prime}(0)|$ is sufficiently small, then the singularity is censored.
\end{prop}

\noindent{\bf Proof:}
We know that $dt_{rad}/dr>0$ for $r>0$ and that a RNG cannot emanate into the trapped region. Thus a necessary condition for the existence of a naked singularity is that $t_H(r)$ is increasing, i.e. $t_H^\prime(r)>0$, on some neighbourhood $(0,a)$ of the origin. This is equivalent to $t_H(r)>t_*(r)$ on $(0,a)$, which in turn is equivalent to
\[ u(r):=m_1-\frac13 m_2-\frac23 r^3(\mot+m_1)^{3/2}(\mo+m_2)>0, \qquad r\in(0,a).\]
Subject to the current hypotheses, we find that
\[ \lim_{r\to 0^+}\frac{u(r)}{r^3}=-2(\mot)^{5/2}-\frac{1}{36}m_2^{\prime\prime\prime},\]
and so the conclusion follows $\bullet$

The following corollary is immediate:

\begin{corr}
If $m_2^\prime(0)=m_2^{\prime\prime}(0)=m_2^{\prime\prime\prime}(0) =0$, then the singularity is censored.
\end{corr}
This is a very useful result as it shows that there is only one case left to consider,
that where $m_2^\prime(0)=m_2^{\prime\prime}(0)=0$ but $m_2^{\prime\prime\prime}(0) \neq 0$.
Thus the structure of the mass function is completely decided for the consideration of this case,
which we call the {\em final case}.

\subsection{The final case}

We assume for the remainder of this section that
$m_2^\prime(0)=m_2^{\prime\prime}(0)=0$ and
$m_2^{\prime\prime\prime}(0) < 0$. The sign of the last term here
is required by the no-shell crossing condition. It is convenient
to define the number $\beta>0$ by
\[ m_2^{\prime\prime\prime}(0)=-12\beta(\mot)^{5/2}.\]
Then from Proposition \ref{hello}, we have, for $r\to 0^+$,
\begin{equation}
m_1(r)\sim-\frac13\beta(\mot)^{5/2}r^3,\qquad
m_2(r)\sim-2\beta(\mot)^{5/2}r^3.\label{mbeq}
\end{equation}
Furthermore, from the proof of Proposition 4, we see that we can
restrict our attention to $\beta\geq 6$; if $\beta<6$, then the
singularity is censored. We will now show that with our approach
one can also easily recover the known result \cite{dj97,jd93}
that, in the final case, sufficient conditions for the existence
of naked singularities are obtained through the solutions of an
algebraic quartic equation. Again the proof follows by calculating
the slopes of the curves in Figure \ref{figure1}.

\begin{prop}
\label{prop-suf}
Subject to the conditions (\ref{mbeq}) which define the final
case, sufficient conditions for the singularity to be naked are
that $\beta>6$ and that there exists at least two positive roots
$q\in(0,\infty)$ of the equation
\begin{equation}
2q(\frac{\beta-q}{6})^{1/3}=3\beta-q. \label{rooteq}
\end{equation}
\end{prop}

\noindent{\bf Proof:} Let $q>0$ and define $t_q(r) = t_0 +
\frac{q}{9}(\mot)r^3$. Notice that $t_q(r)$ intersects the
singularity at $r=0$. Then the region $\{(t,r):0<r<a,
t_q(r)<t<t_H(r)\}$ is non-empty for some $a>0$ if
\[ \beta > q+6>6.\]
We calculate that the slope of a radial null geodesic (RNG)
crossing $t_q$ is
\[ R_{,r}|_{t=t_q(r)} =
(\mot)(\frac{\beta-q}{6})^{-1/3}(\frac{3\beta-q}{6})r^2 +
o(r^2),\] while the slope of $t_q$ is
\[ t_q^\prime(r)=\frac{q}{3}(\mot)r^2.\]
Thus the former slope is less than the latter, and so the RNG
cannot move below $t_q$ as $r$ decreases (hence giving a naked
singularity), provided
\[ (\frac{\beta-q}{6})^{-1/3}(\frac{3\beta-q}{6})<\frac{q}{3}.\]
The result follows $\bullet$

The converse of this result is also true, but is considerably more
difficult to prove. We recall that in past works
\cite{jj-israel,dj97,JJS96} either the emphasis was given on
obtaining only sufficient conditions for the occurrence of a naked
singularity, or the tangent to a null geodesic emanating from the
singularity was assumed to have a finite, positive limit in a
particular co-ordinate system. We note that, in principle, the
non-existence of such limit {\em might not be} a sufficient
condition for the non-existence of a geodesic emanating from the
singularity. We give an example of this in the last section. We
show next a sketch of the rigorous proof that the sufficient
conditions given by Proposition \ref{prop-suf} for the existence
of a naked singularity are in fact also necessary. The details of
the proof are omitted. Unfortunately, the usual fixed point
theorems do not apply, and so an {\em ad hoc} approach is
required.

\begin{prop}
Subject to the conditions (\ref{mbeq}) which define the final
case, a necessary condition for the singularity to be naked is
that there exists a root $q\in[0,\infty)$ of the equation
\[
\label{quartic}
2q(\frac{\beta-q}{6})^{1/3}=3\beta-q.
\]
\end{prop}

\noindent{\bf Sketch of proof:} In this final case, the governing
equation (\ref{rngeq}) for the geodesics may be written as
\begin{equation}
\frac{dy}{du}=(\frac{\beta}{6})^{-1/3}
(1+f_1(u))(1-\frac32\frac{y}{u}+f_2(u))^{-1/3}(1-\frac12\frac{y}{u}+f_3(u)),
\label{yeq}
\end{equation}
where $y=6\beta^{-1}(\mot)^{-1}(t-t_0)$, $u=r^3$ and $f_i(u),
i=1,2,3$ are $C^0$ functions of $u$ which vanish in the limit $u\to
0^+$. We assume that there exists a solution $y_1(u)$ of
(\ref{yeq}); i.e. $y_1\in C^1(0,u_i]$ for some $u_i$, $y_1(0)=0$
and $y_1$ satisfies (\ref{yeq}) on $(0,u_i]$. Furthermore, the
solution must satisfy, for $u\in(0,u_i]$,
\[ 0<y_1(u)<y_H(u)=\alpha u+f_4(u),\]
where $\alpha=(\beta-6)/\beta$, $y_H$ corresponds to the apparent
horizon, and this defines the continuous $o(u)$ function $f_4(u)$.

A key point of the proof, which is used repeatedly, is that {\em
if} the limit $\lim_{u\to 0^+} u^{-1}y_1$ exists, then using
l'Hopital's rule and (\ref{yeq}), this limit must satisfy a
certain quartic equation which strongly restricts the allowed
values of this limit. In fact we {\em do not need} to prove that
this limit exists as the quartic equation arises elsewhere in the
analysis. Furthermore, the existence of this limit is {\em not}
guaranteed by the proof below, and the limit possibly may not
exist.

The proof proceeds by iteratively improving the bounds
$0<y_1/u<y_H/u$ by using existing bounds in (\ref{yeq}) and
integrating. Unfortunately this is not a straightforward process,
as the bounding intervals do not automatically contract at each
step. However, by studying carefully the subregions of a current
bounding interval where contraction does not occur, we can show
that the resulting `expansion' leads to a contradiction such as
$y_1/u>\alpha$. The first step is to show that there exists
$u_0>0$ such that
\[ k_0<\frac{y_1}{u}<l_0 \]
on $(0,u_0]$, where $k_0,l_0$ are respectively the smaller and
greater positive roots of $g_0(x)=x^4-x^3+\lambda_0^3$, where
\[ \lambda_0^3=(\frac{6}{\beta})(\frac{\beta+3}{\beta})^3.\]
Such roots must exist in order that $y_1$ exists. Next, we define
the sequences $\{\lambda_n\}_{n=0}^\infty$, $\{k_n\}_{n=0}^\infty$
and $\{l_n\}_{n=0}^\infty$ as follows. With the $n=0$ terms as
above, for $n\geq 0$ we take
\[
\lambda_{n+1}=\frac32(\frac{6}{\beta})^{1/3}(1-\frac{l_n}{3}),\]
and define $k_n,l_n$ to be the smaller and larger positive roots
of $g_{n+1}(x)=x^4-x^3+\lambda_{n+1}^3$. As for the first step, we
can force the contraction $u^{-1}y_1\in(k_n,l_n)$. Clearly,
$k_{n+1},l_{n+1}$ are not defined for all $n$ for particular
values of $\beta$. When this happens, $u^{-1}y_1$ leaps out of the
current interval $(k_n,l_n)$ giving a contradiction and so a
solution cannot exist. So we assume that $\beta$ is such that the
three sequences are defined. Then it is readily shown that
$\{\lambda_n\}_{n=0}^\infty$, $\{k_n\}_{n=0}^\infty$ are strictly
increasing, while $\{l_n\}_{n=0}^\infty$ is strictly decreasing.
If $\{\lambda_n\}_{n=0}^\infty$ is unbounded, we obtain a
contradiction. Hence this sequence, and consequently
$\{k_n\}_{n=0}^\infty$, $\{l_n\}_{n=0}^\infty$ must converge. It
is easily demonstrated that the limit $l:=\lim_{n\to\infty}l_n$
must be positive and must satisfy
\[
l(1-l)^{1/3}=\frac{3}{2}(\frac{6}{\beta})^{1/3}(1-\frac{l}{3}).\]
Taking $q=\beta l$ proves the result$\bullet$

We note that the proof avoids using the assumption that
$\lim_{u\to0^+}u^{-1}y_1$ exists, nor does it conclude that this
is the case (i.e. we do {\em not} have
$\lim_{n\to\infty}k_n\neq\lim_{n\to\infty}l_n$). The necessary and
sufficient conditions for the occurrence of naked singularities in
the final case can be summarised in the next result which is
obtained by carefully studying all the possible roots of
(\ref{rooteq}):

\begin{corr} In the final case, excluding $\beta=\beta_*:=\frac32(26+15\sqrt{3})\simeq 77.97$, the singularity is naked if and
only if $\beta >\beta_*$.
\end{corr}

\noindent{\bf Proof:} We will see below that for $\beta=\beta_*$,
the equation (\ref{rooteq}) has exactly one positive root. This
does not guarantee that the sufficient condition of Proposition 5
is verified, and so we exclude this case from our considerations.
Then as we have seen, the necessary and sufficient conditions for
the singularity to be naked are that $\beta>6$ and that there
exist at least two positive roots of
\[ f_\lambda(l)=l^4-l^3 + \lambda(1-\frac{l}{3})^3 = 0,\]
where $\lambda = 81/(4\beta)$. The values of $\lambda$ for which
$f_\lambda$ has roots are determined as follows. First, we look
for double roots of $f_\lambda$, i.e. values of $l$ for which
$f_\lambda(l)=f_\lambda^\prime(l)=0$. These occur for two values
of $\lambda$ only; $\lambda=\lambda_1=27(26-15\sqrt{3})/2$ with
double root $l_1=3(2-\sqrt{3})$ and
$\lambda=\lambda_2=27(26+15\sqrt{3})/2$ with double root
$l_2=3(2+\sqrt{3})$. Notice that since $l_1<3$, $f_\lambda(l_1)<0$
for all $\lambda<\lambda_1$, and since $l_2>3$, $f_\lambda(l_2)<0$
for all $\lambda>\lambda_2$. Thus $f_\lambda$ has positive roots
for $\lambda\in(0,\lambda_1)\cup(\lambda_2,\infty)$.

Suppose next that $f_\lambda$ has two positive roots for $\lambda\in(\lambda_1,\lambda_2)$ Then
$f^\prime_\lambda$ has a positive root $l_*$ which can be shown to lie in $(l_1,l_2)$. But the conditions
$\lambda\in(\lambda_1,\lambda_2)$, $f^\prime(l_*)=0$, $l_*\in(l_1,l_2)$ yield $f_\lambda(l_*)>0$, which cannot be true.
Hence $f_\lambda$ cannot have two positive roots for $\lambda\in(\lambda_1,\lambda_2)$.

Thus $f_\lambda$ has (at least) two positive roots iff
$\lambda\in(0,\lambda_1)\cup(\lambda_2,\infty)$. In terms of
$\beta$, this is iff
$\beta\in(0,3(26-15\sqrt{3})/2)\cup(3(26+15\sqrt{3})/2,\infty)$.
The condition $\beta>6$ restricts us to the latter interval
$\bullet$

For convenience, we summarise the results of this section as
follows, phrasing our results in terms of the behaviour of the
initial density function $\mu(r)$.

\begin{theorem}
Let $b>0$ and $\mu\in C^3[0,b]$.  Let
\[ m(r)=\int_0^r x^2\mu(x)dx\]
defined on $[0,b]$ satisfy the no-shell crossing condition
(\ref{nosc}). Then the marginally bound collapse of the dust
sphere with initial radius $b$ and initial density profile
$\mu(r)$ results in a naked singularity if and only if one of the
following conditions is satisfied.
\begin{enumerate}
\item $\mu^\prime(0)<0.$
\item $\mu^\prime(0)= 0$ and $\mu^{\prime\prime}(0)<0$.
\item $\mu^\prime(0)=\mu^{\prime\prime}(0)=0$ and
\[
\frac{\mu^{\prime\prime\prime}(0)}{(\mu(0))^{5/2}}<-\frac23(26\sqrt{3}+45)\simeq-60.0222.\]
\end{enumerate}
\end{theorem}

We note that the results of this section are in complete agreement
with those of previous studies \cite{JJS96,jd93,JS95}. Our
approach, which will be also used in next section, is
substantially different and so gives independent verification of
these past results on radial null geodesics. Furthermore, our
treatment of the final case is more general than it was previously
done since we establish rigorously that, from the RNG's analysis,
the sufficient conditions for the occurrence of naked
singularities are also necessary. We will comment on this again in
the final section.

Having determined fully and rigorously which data give rise
to singularities which are radially visible, we may now consider the case of
non-radial null geodesics.

\section{Non-radial null geodesics}
In this section, we determine the circumstances under which a
non-radial null geodesic may emanate from the singularity. This is
potentially important since, physically, one can expect null
geodesics to have some angular momentum. Furthermore, by
introducing angular momentum in the geodesics equations we will
also be doing a stability study of the naked singularity solutions
that appear in the radial case.

The governing equations for the non-radial null geodesics (NRNG's) obtained
from the Euler-Lagrange equations are
\begin{eqnarray}
{\ddot t}+\frac{R_{,rt}}{R_{,r}}{\dot
t}^2+\frac{L^2}{R^2}(\frac{R_{,t}}{R}-\frac{R_{,rt}}{R_{,r}})&=&0,
\label{NRNG1}\\
{R_{,r}}{\ddot r}+R_{,rr}{\dot r}^2+2R_{,rt}{\dot r}{\dot
t}-\frac{L^2}{R^3}&=&0,\label{NRNG2}\\
-{\dot t}^2+(R_{,r}{\dot r})^2+\frac{L^2}{R^2}&=&0. \label{NRNG3}
\end{eqnarray}
The over-dot represents differentiation with respect to an affine parameter $s$ along the
geodesics, and the constant $L$ is the conserved angular momentum. We are looking for is
the existence of a solution of
(\ref{NRNG1}-\ref{NRNG3}), which satisfies this
\\\\
{\bf Existence condition:}
\newline
There exists $\epsilon>0$ such that ${\dot t}$ is a non-negative,
integrable function of the affine parameter $s$ and ${\dot r}$ is
an integrable function of $s$ for $s\in[0,\epsilon)$ and such that
\[ \lim_{s\to 0^+} r(s) =0,\qquad \lim_{s\to 0^+} t(s) = t_0.\]

Non-negativity of ${\dot t}$ implies that the geodesic is
future-pointing. We immediately see from (\ref{NRNG3}) that ${\dot
t}$ must be infinite in the limit $s\to0^+$, and hence we must
have ${\ddot t}\to-\infty$ as $s\to 0^+$. In fact this applies to
any NRNG which extends back to the center $R=0$ ($r=0$).
These conditions will be used in the proofs that follow. The
main idea of the proofs given in this section is
essentially the same as in the case of RNG where in order to have
naked singularity solutions, one needed to find geodesics $\gamma$
which lie in the region between $t_H$ and $t_*$ (as shown in
Figure \ref{figure1}). The existence of the lower bound $t_*$
obtained in the next proposition is a simple consequence of the
geodesic equations (\ref{NRNG1}-\ref{NRNG3}).

\begin{prop}
A non-radial null geodesic emanating from the singularity must lie
in the region $t_*(r)<t\leq t_H(r)$ for all $r\in(0,a)$ and some
$a>0$.
\end{prop}

\noindent{\bf Proof:} The second inequality is obvious. For the
first, note that using the geodesic and field equations we can
write
\begin{equation}
\label{ddot}
{\ddot t}=-\frac{R_{,rt}}{R_{,r}}(R_{,r}{\dot
r})^2+\frac{L^2}{R^2}(\frac{m}{R^3})^{1/2},
\end{equation}
and so
\begin{equation}
{\ddot t}\geq-\frac{R_{,rt}}{R_{,r}}(R_{,r}{\dot
r})^2.
\end{equation}
Thus for a solution which extends back to the singularity, we must
have $R_{,rt}>0$, which gives $t>t_*(r)$ (recall that this is the
defining property of $t_*(r)$) $\bullet$

We can now consider the following problem: Suppose that for a
given mass function $m(r)$ there are no RNG's escaping from the
singularity. Can we find NRNG's which escape? Before giving the
answer below in Proposition \ref{ola} we need to prove the
following result:

\begin{lemma}\label{lemma1}
${\dot r}\neq 0$ in a neighbourhood of $s=0^+$ along an NRNG which
extends back to the centre $r=0$.
\end{lemma}

\noindent{\bf Proof:} Suppose that ${\dot r}=0$ at some point
$s=s_0$ on such a geodesic. From (\ref{ddot}) this gives $\ddot t$
positive which cannot happen sufficiently close to $r=0$ $\bullet$

Note that this lemma applies to the case where the geodesic
extends back to the singularity. We can now give an answer to the previous question
by simply using the geodesics equations (\ref{NRNG1}-\ref{NRNG3}) and
the results of the last section on RNGs:

\begin{prop}
\label{ola}
If the singularity is radially censored, then it is censored. That
is, if there are no radial null geodesics emanating from the
singularity, then there are no null geodesics emanating from the
singularity.
\end{prop}

\noindent{\bf Proof:} Assume that the singularity is radially
censored and suppose that there exists an NRNG which extends back
to the centre. By Lemma 1, we have ${\dot r}\neq0$ and since
generally $R_{,r}\neq0$, we can rewrite (\ref{NRNG3}) as
\[(\frac{{\dot t}}{R_{,r}{\dot r}})^2=1+(\frac{L}{RR_{,r}{\dot
r}})^2>1.\] Thus along an NRNG, we have
\[\frac{dt}{dr}>R_{,r}.\]
The right hand side here gives the slope of an outgoing radial
null geodesic, the left hand side is the slope of the NRNG's. Now
let $p$ be an arbitrary point of $\Omega=\{(t,r):r>0,t_*(r)<t<t_H(r)\}$. There is a
unique outgoing radial null geodesic through $p$ which by
hypothesis extends back to the regular centre rather than the
singularity. The inequality above tells us that {\em any} NRNG
through $p$ must cross the radial null geodesic from above, and so
must precede this geodesic at all points $q$ preceding $p$. Hence
no NRNG's can extend back to the singularity $\bullet$

Now we consider the converse problem: Suppose that for a given
mass function $m(r)$ there are RNG's escaping from the
singularity. Will there necessarily be NRNGs which also escape? To
answer this problem we found it convenient to define new
variables:
\[ x = {\dot t},\qquad y=R_{,r}{\dot r}.\]
Then a crucial point will be with how to control the derivative of
$x/y$ as $r\to 0$. We explain why this should be done before we
actually prove our result in Proposition \ref{oi} ahead.

We assume henceforth that there
exist radial null geodesics emanating from the singularity.
Choosing
data at a point $p$ with $x|_p>0$, $y|_p>0$, which corresponds to a future
pointing outgoing null geodesic, we see that $(x,y)$ lies on the
positive right-hand branch of the hyperbola ${\cal H}_L$ given by
\begin{equation} x^2-y^2=\frac{L^2}{R^2}. \label{hyp} \end{equation}
Clearly $x/y>1$ along this branch. This says that the slope
$dt/dr$ in the $r-t$ plane exceeds the slope of the unique
future-pointing outgoing RNG through $p$. However, since $p$ is an
interior point of $\Omega$, we can find $\delta$ sufficiently
small such that the unique solution through $p$ of $x/y =
1+\delta$ will also extend back to the singularity. (Notice that
the proofs of existence of naked singularities for the radial case
all involved showing that the slope $dt/dr$ was smaller that the
slope of some curve which we know to intersect the singularity.)
Now on the hyperbola ${\cal H}_L$, there is a continuum of choices
of $(x,y)$ which satisfy $x/y <1+\delta$. Take any such pair and
let this give the initial data for (\ref{NRNG1}-\ref{NRNG3}) at
some time $s_0>0$. We follow the NRNG with these data as $s$
decreases. Letting $s\to 0^+$, we must have $x/y >1$ for all
$s>0$, which says that $\gamma$ stays {\em below} the RNG through
$p$. By our choice of data,
\[ (\frac{x}{y})(s_0)<1+\delta;\]
to prove that $\gamma$ extends back to the singularity, it would
suffice to show that
\[ (\frac{x}{y})(s)<1+\delta, \qquad s\in(0,s_0].\]
This inequality says that $\gamma$ stays {\em above} the curve
with $x/y = 1+\delta$, which we know extends back to the
singularity. This would prove that the NRNG must also extend back
to the singularity. The key to proving that we may proceed in this
manner is to identify a subset of $\Omega$ wherein the rate of
change of $x/y$ may be controlled. This is done by carefully
analysing the relative magnitudes of the terms that arise in the
equation governing this derivative as will be shown in the next
proposition. (Note however that the demonstration of the existence
of NRNG's emanating from the singularity does not quite follow the
outline of the paragraph above: this paragraph is intended to
motivate our attempt to control the derivative of $x/y$.)

\begin{prop}
\label{oi} Let $m(r)$ be such that there exist
radial null geodesics emanating from the singularity.
Let
\[ G(r,t)=R\frac{R_{,rt}}{R_{,r}}.\] For $r_0>0$, $n>0$, $\delta>0$ define
\[\Omega[n,r_0]=\{(r,t): 0<r\leq r_0,\,
G(r,t)>\frac{n-1}{2n}\sqrt{\frac{m}{R}},\, t<t_H(r)\}\] and
\[\Omega[\delta,n,r_0]=\{(r,t):0<r\leq r_0,\,
\sqrt{\frac{m}{R}}>\frac{2n}{3n-1}(1+\delta),\, t<t_H(r)\}.\] Then
there exists $r_0>0$, $n>0$ and $\delta>0$ such that for any
$p\in\Omega[\delta,n,r_0]$, an NRNG with initial data satisfying
$(\frac{x}{y})|_p \leq 1+\delta$ extends back to the singularity.
\end{prop}

\noindent{\bf Proof:} Straightforward calculations show that
$(r,t)\in\Omega[n,r_0]$ if and only if $t_n(r)<t<t_H(r)$ where
\[t_n=\frac{3n}{3n-1}t_c-\frac{2}{3n-1}\frac{\sqrt{rm}}{m_{,r}},\]
and that $(r,t)\in\Omega[\delta,n,r_0]$ iff $t_\delta<t<t_H$ where
\[ t_\delta=t_c-\frac23\sigma^{-1}m\]
and $\sigma^{1/3}=2n(1+\delta)/(3n-1)$.

The inequalities $t_*(r)<t_n(r)<t_\delta(r)<t_H(r)$ may be shown
to be true for sufficiently small $r$. This is easily done except
in the final case, i.e. where
$m_1(r)\sim-\frac13\beta(\mot)^{5/2}r^3$. In this case, the
relevant inequalities will hold if $n$ is chosen so that
\[ \beta > \max\{n_1=3(3n-1)(\frac{3n-1}{2n})^3,
n_2=6(\frac{4n-1}{n-1})(\frac{3n-1}{2n})^3\}\] (In fact, the
condition that $\beta>n_2$ arises from a later consideration, but
we find it convenient to deal with it here). Rather fortuitously,
we find that taking $n_1=n_2$ gives $n_1=n_2=\beta_*$, and since
we must have $\beta>\beta_*$, the inequalities are verified. The
inequality $t_\delta<t_H$ requires that $\delta>0$ be chosen
sufficiently small.

Using the geodesic equations (\ref{NRNG1}-\ref{NRNG3}), we find
that
\begin{equation}
\frac{d}{ds}(\frac{x}{y})
=\frac{L^2}{R^3y}\left(G+\sqrt{\frac{m}{R}}-\frac{x}{y}\right).
\label{xydot} \end{equation} Now let $p$ as defined be the initial
point at time $s=s_0$ of an NRNG and set the data such that
$(\frac{x}{y})(s_0)\leq 1+\delta$. By definitions, while the NRNG
remains in $\Omega[\delta,n,r_0]$, we have
$\frac{d}{ds}(\frac{x}{y})>0$. As $s$ decreases, this geodesic
cannot leave $\Omega[\delta,n,r_0]$ through $t=t_H$ (this would
correspond to a future-pointing null geodesic exiting the trapped
region across its past space-like boundary). Suppose that the
geodesic reaches the boundary $t=t_\delta(r)$. In the $r-t$ plane,
the geodesic has slope $t^\prime = \frac{x}{y}R_{,r}$. But for
$s\leq s_0$, while the geodesic is in the closure of
$\Omega[\delta,n,r_0]$, we have (since $x/y$ decreases as $s$
decreases)
\[ t^\prime\leq(1+\delta)R_{,r}.\]
Also, $(1+\delta)R_{,r}|_{t=t_\delta(r)}<t_\delta^\prime(r)$ for
sufficiently small $r$ and $\delta$. (As above, this is immediate
except in the final case, where is necessitates the inequality
$\beta>n_2$.) Thus the geodesic cannot exit $\Omega[\delta,n,r_0]$
via this boundary, and therefore must extend back to the
singularity $\bullet$

This then proves that if, for a given mass function $m(r)$, RNGs
escape from the singularity then NRNGs will also necessarily
escape. The following theorem summarises the principal results of
this section.

\begin{theorem}
Given regular initial data for marginally bound spherical dust
collapse subject to a condition which rules out shell-crossing
singularities, there are non-radial null geodesics which emanate
from the ensuing singularity if and only if there are radial null
geodesics which emanate from the singularity.
\end{theorem}
Finally, it is also of interest to note that there are always NRNG's which,
starting from any point $p$ with $t<t_H(r)$, avoid the singularity
in the past:

\begin{prop}
Let $p$ lie in the region $\{(r,t):r>0,0<t<t_H(r)\}$. If the initial data
for an NRNG $\gamma$ through $p$ satisfy $(\frac{x}{y})|_p
>\frac32$, then $\gamma$ cannot meet the singularity in the past.
\end{prop}

\noindent{\bf Proof:} For $t_*<t<t_H$, it is easily shown that
\begin{equation}
 G(r,t) <\frac12\sqrt{\frac{m}{R}}.\label{gbound}
 \end{equation} Suppose that $\gamma$
meets the singularity. Then from above, it must do so in the
region $t_*<t<t_H$ and the condition that $0>{\dot x}={\ddot
t}\to-\infty$ in a neighbourhood of the singularity yields, using
(\ref{NRNG1}),
\[ (\frac{x}{y})^2 <\frac32.\]
But (\ref{xydot}), (\ref{gbound}) and the initial condition at $p$
yield $\frac{d}{ds}(\frac{x}{y})|_p <0$. Hence the condition $x/y > 3/2$
is maintained along the geodesic moving into the past, and so we
obtain a contradiction. Hence $\gamma$ cannot meet the singularity
in the past$\bullet$

Having proved the main results of this paper we will
discuss some of their
consequences in the next section.
\section{Conclusions and comments}
We have shown that the existence of radial null geodesics
emanating from the shell-focussing singularity in marginally bound
spherical dust collapse is a necessary and sufficient condition
for the existence of non-radial null geodesics emanating from the
singularity. Our interpretation of this result is that it
expresses an aspect of stability of the naked singularity: the
insertion of angular momentum into the system, at the level of the
null geodesics, does not affect the outcome. This result can be
interpreted as implying the robustness of the previous results,
for radial null geodesics, with respect to the presence of angular
momentum. One consequence of this is that all such naked
singularities are equally `malicious': previously, it has been
noted that those which arise by the conditions (i) or (ii) of
Theorem 1 being satisfied are gravitationally weak, while those
which arise by condition (iii) of this theorem being satisfied are
gravitationally strong \cite{JJS96}. However it is known that a
non-radial null geodesic emanating from a central singularity {\em
always} satisfies the strong singularity condition
\cite{nolan2000}, and so the distinction in terms of singularity
strength vanishes when angular momentum is included.

Our results on the relationship between the initial data and the
causal nature of the singularity agree with previous results based
on the study of radial null geodesics. There are some differences:
we have taken the initial data $\mu(r)$ to be $C^3$, as opposed to
analytic or $C^\infty$. The convenience of this assumption is that
it is more general and allows a comprehensive and rigorous
discussion of all possible cases. Can this level of
differentiability be reduced? Some authors have studied this
problem (in fact the more general problem where the collapse is
not assumed to be marginally bound) using $C^1$ initial data (e.g.
\cite{jj-israel,sary-ghate}; here a differentiability condition is
imposed on the mass function $m(r)$ which corresponds to $\mu(r)$
being $C^1$). However in these studies either a complete
decomposition of the initial data space into regions giving rise
to naked and censored singularities has not been obtained, or
further restrictions on the differentiability are imposed. It
seems to us that the critical role of the third derivative of
$\mu(r)$ indicates that $C^3$ is as low as one can push the
differentiability requirement while maintaining a complete
description of the possible outcomes.

As mentioned above, the approach used here to prove the existence
of null geodesics emanating form the singularity is different from
that of previous studies (but is more in line with the earlier
studies \cite{christo,newman}). One immediate advantage is that we
can give independent verification of the previous results on
radial null geodesics. We see other two distinct advantages to our
(more qualitative) approach. Firstly, we do not need to assume the
existence of a limit of $R/r^\alpha$ (for some $\alpha>0$) in the
limit as the singularity is approached. Assuming the existence of
such a limit, or equivalently of $y_1(u)/u$ as $u\to0^+$ above,
removes the need for the detailed analysis of Proposition 6: one
can simply take the limit of both sides of (\ref{yeq}) and use
l'Hopital's rule to produce the quartic equation of Corollary 4.
However, while it is almost certain that this limit exists for
some of the null geodesics emanating from the singularity, it is
by no means certain that this is the case for {\em all} such
geodesics. Evidence that the limit need not exist arises from the
contraction process in the proof of Proposition 6. If the limit
did exist, one would expect this to be reflected in the bounding
interval $(k_n,l_n)$ having zero length in the limit as
$n\to\infty$. In fact this does not happen, so while $y_1(u)/u$ is
bounded within $[\lim_{n\to\infty}k_n,\lim_{n\to\infty}l_n]$ in
the limit as $u\to0^+$, there is no evidence that it possesses a
finite limit. By way of illustration, we note that the approach
relying on the existence of such a limit would not identify the
(positive) solution \[y(x)=x(1+\epsilon\sin\frac1x)\qquad
(0<\epsilon<1)\] of the initial value problem
\[y^\prime(x)=\frac{y}{x}-
\frac{\epsilon}{x}\sqrt{1-\frac{1}{\epsilon^2}(1-\frac{y}{x})^2},\qquad
y(0)=0.
\]
Of course it may be that the regularity conditions imposed on the
initial data only allow geodesics for which such a limit exists;
however we feel that this is something which needs to be checked.
Our fundamental point here is that the non-existence of this limit
might not be a sufficient condition for the absence of a geodesic
emanating from the singularity.

A second advantage which we see is that the qualitative approach
applied here lends itself very easily to the analysis of
non-radial null geodesics. In future work we hope to show that
this is also the case when one considers radial and non-radial
time-like geodesics. This would then give a unified treatment of
all causal geodesics in marginally bound spherical dust collapse.
One clear disadvantage must be mentioned: We have dealt only with
marginally bound, and not general, spherical dust collapse. We
expect that the analysis used here can be carried over to the
general case, but it must be acknowledged that this would not be
done without some difficulty. In the general case, one no longer
has the explicit solution (\ref{req}) and consequently,
identifying the important surfaces such as $t=t_*(r)$ becomes
problematic.

We comment as follows on the relationship between the initial data
and the temporal nature of the singularity. One can argue that
$\mu^\prime(0)\neq 0$, i.e. $m_1^\prime(0)\neq0$ is unphysical, as
it represents a cusp in the initial density distribution. One can
also argue that all higher order odd derivatives should vanish at
$r=0$, i.e. that one should take $\mu(r)$ to be an even function
of $r$ \cite{christo,newman}. The reasoning is that the polar co-ordinates
$(r,\theta,\phi)$ are singular at $r=0$, and so one should move to
a Cartesian co-ordinate system $(x,y,z)$ in a neighbourhood of the
origin. But then odd powers of $r$ lead to non-smooth functions of
$(x,y,z)$. In particular, demanding $C^2$ behaviour in the
Cartesian co-ordinate system (it is difficult to see where higher
derivatives would play a role in the physics, e.g. particle
motion, of the space-time on and shortly after the initial slice)
would rule out an $r^3$ term in $\mu(r)$ and enforce
$\mu^{\prime\prime\prime}(0)=0$. This leaves
$\mu^{\prime\prime}(0)$ as the only feature of regular initial
data which has any input into determining whether the collapse
results in a naked or censored singularity. Clearly, the generic
case is $\mu^{\prime\prime}(0)<0$ which leads to a naked
singularity.

Allowing the rather mild singularities generated by non-vanishing
$\mu^\prime(0)$ or $\mu^{\prime\prime\prime}(0)$, the picture
becomes less clear. Obviously naked singularities will arise as
the generic case, but a more interesting question remains open: Do
small perturbations of vacuum $\mu\equiv 0$ and homogeneous
$\mu\equiv\mu(0)$ initial data generically give rise to naked
singularities? Consider, for example, the following situation.

 Fix  the $C^3$ function $m_1(r)$,
satisfying $m_1^\prime(0)=m_1^{\prime\prime}(0)=0$. Then we have,
as in Section 3, $m_1(r)\sim
-\frac13\beta(\frac{\mu_0}{3})^{5/2}r^3$. Set initial data
$(\mu_0,m_1(r))$ and allow the collapse to proceed. Taking
$\beta<\beta_*$, the singularity is censored. Now consider the
initial data $({\bar\mu}_0,{\bar m}_1(r)=m_1(r))$. Then
\[{\bar\beta} = (\frac{\mu_0}{ {\bar\mu}_0 })^{5/2}\beta. \]
Choosing ${\bar\mu}_0$ sufficiently {\em small}, we can arrange
that ${\bar{\beta}}$ exceeds the numerical value $\beta_*$ and so
the collapse will result in a naked singularity. The singularity
will occur at time
\[{\bar t}_0 =
\frac23(\frac{{\bar\mu}_0}{3})^{-1/2}>t_0=\frac23(\frac{\mu_0}{3})^{-1/2}.
\]
Thus there is a tendency for `smaller' initial data to result in
naked rather than censored singularities. We note that this
pattern reflects what is frequently found in studies of critical
phenomenon: for example, in spherical collapse of a massless
scalar field, Choptuik space-time, which contains a naked
singularity, lies on the boundary between small/weak initial data
which lead to dispersal, and  large/strong initial data which lead
to black hole formation \cite{carsten}. It would be useful to
carefully analyse this issue as it arises in dust using a standard
measure of initial data for general relativity \cite{he,klain}.

\subsection*{Acknowledgements} BCN acknowledges support from DCU under the Albert
College Fellowship Scheme 2001, and hospitality from the
Relativity Group at Queen Mary, University of London. FCM thanks
Reza Tavakol and Malcolm MacCallum for discussions, CMAT, Univ.
Minho, for support and FCT (Portugal) for grant PRAXIS XXI
BD/16012/98.

\end{document}